\documentclass[lettersize,journal]{IEEEtran}

\usepackage{amsmath,amsfonts}
\usepackage{algorithmic}
\usepackage{algorithm}
\usepackage{array}
\usepackage[caption=false,font=normalsize,labelfont=sf,textfont=sf]{subfig}
\usepackage{textcomp}
\usepackage{booktabs}
\usepackage{graphicx}
\usepackage{stfloats}
\usepackage{url}
\usepackage{verbatim}
\usepackage{cite}

\usepackage{tikz}
\usetikzlibrary{
    shapes.geometric,
    shapes.symbols,
    arrows.meta,
    positioning,
    calc,
    decorations.pathmorphing,
    patterns
}
\usepackage{pgfplots}
\pgfplotsset{compat=1.17}

\pgfdeclarelayer{background}
\pgfsetlayers{background,main}

\usepackage[hidelinks]{hyperref}

\begin{document}

\title{ActiveFlowMark: Assessing Tor Anonymity under Active Bandwidth Watermarking}

\author{
    \IEEEauthorblockN{
    Zilve Fan\IEEEauthorrefmark{1},
    Zijian Zhang\IEEEauthorrefmark{1},
    Yangnan Guo\IEEEauthorrefmark{1},
    Jiaqi Gao\IEEEauthorrefmark{1},
    Zhen Li\IEEEauthorrefmark{1},
    Mengyu Wang\IEEEauthorrefmark{2},
    Chengxiang Si\IEEEauthorrefmark{2},
    and Liehuang Zhu\IEEEauthorrefmark{1} 
    }
  
\thanks{Corresponding author: Zijian Zhang}
\thanks{Z. Fan, Z. Zhang, Y. Guo, J. Gao, Z. Li, and L. Zhu are with the School of Cyberspace Science and Technology, Beijing Institute of Technology, Beijing 100081, China (E-mail: \{3220235140, zhangzijian, guoyangnan, gaojiaqi, zhen.li, liehuangz\}@bit.edu.cn).}
\thanks{M. Wang is with the School of Cyberspace Science and Technology,
Beijing Institute of Technology, Beijing 100081, China, and also with the National
Computer Network Emergency Response Technical Team/Coordination Center
of China, Beijing 100029, China (E-mail: wangmengyu@cert.org.cn).}
\thanks{C. Si is with the National Computer Network Emergency Response Technical Team/Coordination Center of China (CNCERT/CC), Beijing 100029, China (E-mail: sichengxiang@cert.org.cn).}
}

\maketitle

\begin{abstract}
Low-latency anonymity networks such as Tor remain vulnerable to infrastructure-level traffic analysis that exploits side-channel information observable from encrypted communications. 
We introduce NATA, a non-invasive active traffic-correlation analysis algorithm that injects distinguishable throughput patterns into traffic flows through controlled bandwidth perturbations. 
Unlike passive correlation methods, NATA does not require endpoint compromise, Tor-browser modification, or packet-payload decryption or modification. 
It can be carried out by an adversary that controls an upstream network gateway and observes traffic at adversary-controlled exit relays.

To identify perturbed flows under substantial network variability, we develop BM-Net (Bandwidth Modulation Network), a selective state-space learning framework adapted for bandwidth-modulation detection. 
Given the limited availability of high-fidelity ground truth on real-world cross-continental Tor paths, BM-Net adopts a data-efficient learning strategy that separates self-supervised representation learning from supervised task-specific classification.
It first learns reusable traffic representations through masked pre-training on serialized traffic traces, and then adapts these representations to binary perturbation detection and fine-grained modulation classification using task-specific labeled data.

Through real Tor traffic measurements, BM-Net achieves a 99.65\% binary detection F1 score and a 97.5\% macro-F1 score for fine-grained modulation classification under our evaluated settings. 
In addition, \textit{tornettools}-based scaled simulations are used to estimate exit-observation probability under bandwidth-weighted relay selection. 
These results suggest that active bandwidth perturbation can serve as an infrastructure-level side channel for traffic correlation under a clearly defined adversary model.
\end{abstract}


\section{Introduction}
\IEEEPARstart{T}{or} is one of the most widely deployed low-latency anonymity networks. 
It routes user traffic through a sequence of relays with layered encryption to protect the relationship between users and their destinations. 
Despite these protocol-level protections, Tor remains vulnerable to traffic-analysis attacks that exploit metadata observable from encrypted flows, thereby weakening its anonymity guarantees.

Most existing traffic-analysis attacks follow a passive correlation paradigm, in which an adversary infers relationships between flows from naturally occurring timing, volume, or direction patterns. 
However, the reliability of passive correlation can be reduced by traffic obfuscation mechanisms, pluggable transports such as Obfs4, and background chaff traffic, all of which may distort or mask the statistical features used by passive detectors. 
In such settings, passive evidence may remain probabilistic and can be sensitive to network noise, traffic load, and defense configurations.

In this paper, we study an active traffic-correlation setting in which an infrastructure-level adversary intentionally perturbs the bandwidth of selected Tor traffic. 
Rather than relying only on incidental metadata leakage, the adversary introduces controlled throughput variations that can later be observed at adversary-controlled exit relays. 
We formalize this approach as Non-invasive Active Traffic-correlation Analysis (NATA). 
The core idea is active flow watermarking through throughput dynamics: a gateway-level shaper imposes low-frequency bandwidth patterns, such as square, sinusoidal, or triangular waveforms, on Tor traffic. 
These patterns create distinguishable throughput signatures that can increase traffic-correlation risk under the considered threat model. 
Because the perturbation is imposed through bandwidth constraints, logical-layer padding alone may not fully remove the resulting throughput pattern in the evaluated settings.

Applying active bandwidth perturbation to Tor introduces several practical constraints. 
First, Tor is a low-latency anonymity network, so bandwidth shaping must avoid excessive degradation that would cause user suspicion, connection stalls, or circuit replacement. 
Second, path variability, relay congestion, and cross-region network jitter can distort the injected waveform before it is observed at the exit side. 
Third, Tor's end-to-end flow-control mechanism, including SENDME cells, may interact with aggressive shaping and cause circuit stalls or termination. 
These constraints make the attack a trade-off between watermark detectability and connection stability rather than a purely signal-injection problem.

We model traffic correlation as a two-stage problem: detecting whether a flow has been perturbed, and then identifying the corresponding modulation pattern. 
To recover bandwidth signatures from noisy Tor traces, we develop BM-Net, a data-efficient deep learning framework for bandwidth-modulation detection. 
BM-Net separates representation learning from task-specific classification. 
It first learns reusable traffic representations through self-supervised masked pre-training on serialized traffic traces, and then adapts these representations through supervised fine-tuning for binary perturbation detection and fine-grained modulation classification. 
This design is intended to reduce the amount of high-fidelity multi-class training data required for real-world cross-continental Tor measurements.

This paper makes the following contributions:
\begin{itemize}
    \item We propose NATA, a non-invasive active traffic-correlation
method that requires no endpoint compromise, no Tor-browser
modification, and no malicious scripts. Operating at the network
infrastructure layer, such as an ISP or upstream gateway, NATA shows
that bandwidth control from this vantage point can create a measurable
side channel for Tor traffic correlation.

    \item We develop BM-Net, a bandwidth-modulation detection framework built around a selective state-space traffic encoder. 
BM-Net adapts serialized packet-level representations, self-supervised masked pre-training, and supervised fine-tuning to the active bandwidth-perturbation setting. 
This design supports data-efficient adaptation under limited labeled data and dataset shift. 
In our real-world measurements, BM-Net achieves a 97.5\% macro-F1 score on a curated small-sample multi-class dataset.
    
    \item We construct a probabilistic model to estimate the likelihood that watermarked Tor traffic is observed by adversary-controlled exit relays. 
    Using \textit{tornettools} simulations with historical consensus data, we quantify how exit-relay control and bandwidth-weighted path selection affect the probability of observing watermarked flows, providing a network-scale estimate of traffic-correlation risk.
\end{itemize}

The remainder of this paper is organized as follows. 
Section~\ref{sec:related} reviews existing traffic-analysis techniques and places our infrastructure-level active approach within the broader landscape of anonymity analysis. 
Section~\ref{sec:threat} presents the threat model, the Shaper--Sniffer architecture, and the research questions. 
Section~\ref{sec:system_architecture} describes the NATA design, including constraint-aware traffic shaping and the two-stage BM-Net training pipeline. 
Section~\ref{sec:theoretical_modeling} derives the probability model for traffic correlation under network-scale deployment scenarios. 
Section~\ref{sec:evaluation} presents the evaluation, combining \textit{tornettools} simulations with real-world Tor measurements. 
Section~\ref{sec:discussion} discusses limitations, deployment implications, and ethical considerations. 
Section~\ref{sec:conclusion} concludes the paper.

\section{Related Work}
\label{sec:related}

Low-latency anonymity networks such as Tor face traffic-analysis threats at the client, relay, and network-infrastructure levels. 
For comprehensive surveys, we refer readers to Karunanayake et al.~\cite{karunanayake2021deanonymisation} and Basyoni et al.~\cite{basyoni2020traffic}. 
Tor provides practical anonymity against many local adversaries, but it does not eliminate metadata leakage from timing, volume, routing, or relay-level observations~\cite{dingledine2004tor}.

\subsection{Traffic Analysis and Flow Correlation}

Passive traffic analysis infers sensitive information from naturally occurring metadata without modifying target traffic. 
Website fingerprinting has evolved from statistical classifiers~\cite{herrmann2009website,hayes2016kfingerprinting} to deep neural models such as Deep Fingerprinting~\cite{sirinam2018deep}, with subsequent work studying robustness and evaluation pitfalls~\cite{juarez2014critical}. 
Flow correlation instead links ingress and egress flows. 
Early timing-based attacks~\cite{murdoch2005low} were followed by learning-based systems such as DeepCorr~\cite{nasr2018deepcorr}, DeepCoFFEA~\cite{oheun}, FlowTracker~\cite{guan}, and AttCorr~\cite{li}. 
These systems primarily exploit natural timing and volume relationships. 
In contrast, our setting introduces controlled bandwidth perturbations and studies whether the resulting throughput patterns can be recovered after Tor-induced distortion.

\subsection{Active Watermarking, Relay-Level Attacks, and Defenses}

Active traffic-analysis techniques intentionally perturb traffic to facilitate later correlation. 
Prior work modifies inter-packet timing or flow rates to embed recognizable signals~\cite{rainbow,swirl,ibw,icbw}, while other attacks exploit clock skew, latency leakage, congestion effects, padding-cell behavior, relay selection, or relay availability mechanisms~\cite{murdoch2006hot,shmatikov2006timing,rochet,iacovazzi,yang,Biryukov,overlier2006locating,borisov,winter2014spoiled}. 
Our work differs by focusing on bandwidth perturbation by an infrastructure-level shaper and by using deep sequence modeling to recover low-frequency throughput patterns from real Tor measurements.

Defenses such as padding, traffic morphing, adversarial trace generation, and lightweight obfuscation can reduce passive leakage but often introduce bandwidth or latency overhead~\cite{cherubin2017website,wright2009traffic,rahman2020mockingbird,gong2020zero}. 
Infrastructure-aware defenses such as Counter-RAPTOR, CLAPS, LASTor, and AS-aware path selection reduce exposure to network-level adversaries but involve performance and deployability trade-offs~\cite{sun2017counter,rochet2020claps,akhoondi2012lastor,edman2009as}. 
These defenses motivate evaluating whether logical-layer obfuscation is sufficient against active bandwidth constraints.

\section{Threat Model, Motivation, and Research Questions}
\label{sec:threat}

In this section, we define the adversary model considered in this work. 
We focus on an infrastructure-level adversary that can shape Tor-related traffic near the client side and observe traffic at adversary-controlled exit relays. 
We then motivate active throughput shaping as a way to create measurable traffic-correlation signals under noisy network conditions, and finally present the research questions that guide our system design.

\subsection{Threat Model}
\label{subsec:threat_model}

We consider a capable but bounded adversary that aims to increase traffic-correlation capability against Tor users through active bandwidth perturbation. 
The adversary does not compromise the client host, modify the Tor browser, inject scripts, or decrypt packet payloads. 
Instead, the adversary operates from two network vantage points, forming a Shaper--Sniffer architecture:

\begin{itemize}
    \item \textbf{Shaper.} 
    The adversary controls a network gateway, Autonomous System (AS), ISP link, or other upstream network position near the target client. 
    From this position, the adversary can apply rate limits or bandwidth modulation to Tor-related TCP connections. 
    The Tor client is treated as a black box: the adversary does not require access to the host machine, Tor process, or browser state.

    \item \textbf{Sniffer.} 
    The adversary operates one or more Tor exit relays and passively records packet-level observations from flows that traverse these relays. 
    These observations may include timestamps, packet sizes, directions, and byte-level packet representations, but the analysis does not rely on decrypting payloads or inspecting application-layer content.
\end{itemize}

In our evaluated setting, the entry-side shaper identifies Tor-related TCP connections using flow-level metadata, such as the client IP address, relay IP address, port, and transport protocol. 
The shaping is applied to the Tor connection between the client and its entry/guard relay. 
Because Tor may multiplex multiple circuits over a single TCP connection, the attack operates at the connection level rather than assuming direct visibility into individual Tor circuits. 
Non-Tor flows are not intentionally shaped. 
For Tor bridges, pluggable transports, VPNs, or other tunneling settings, additional identification assumptions may be required; we discuss these cases as limitations.

For traffic correlation, the shaper imposes a specific throughput pattern on Tor-related traffic near the client side, while the sniffer observes traffic at adversary-controlled exit relays. 
The adversary then uses a pattern-recognition model to determine whether the exit-side traffic contains the injected throughput pattern. 
A match between the injected pattern and an observed exit-side flow provides evidence for correlating the corresponding client-side and exit-side traffic streams.

\begin{figure*}[h]
    \centering
    \includegraphics[width=0.82\linewidth]{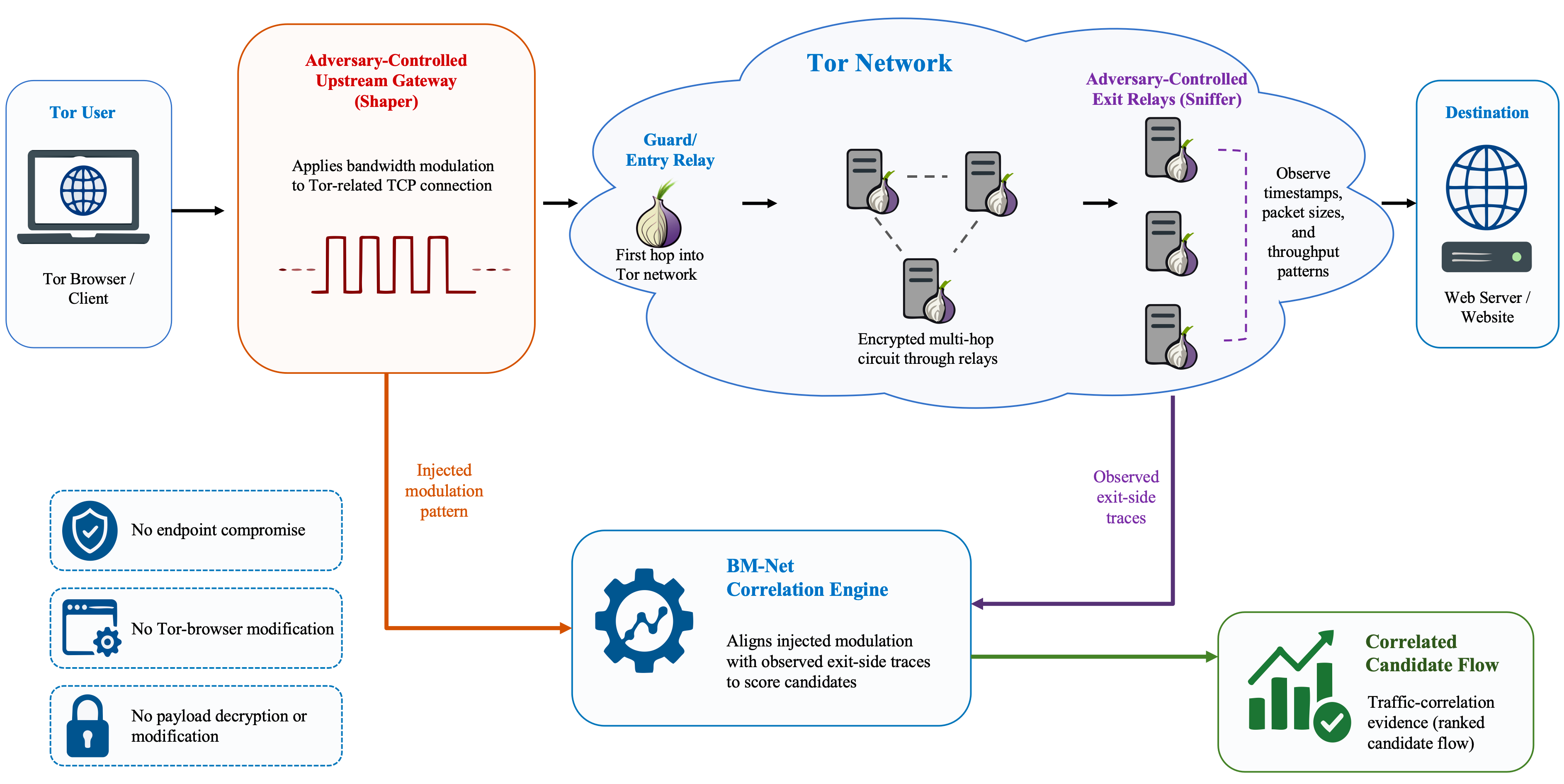}
    \caption{Operational threat model. The adversary shapes Tor-related traffic near the client side and passively observes traffic at adversary-controlled exit relays. The observed traces are analyzed using a hierarchical learning framework to detect and classify bandwidth-modulation patterns.}
    \label{fig:threat_model_diagram}
\end{figure*}

\subsection{Motivation}
\label{subsec:motivation}

Existing traffic analysis against Tor often relies on passive statistical features, such as inter-packet timing, packet-size distributions, and traffic-volume patterns. 
However, these features can be distorted by network jitter, variable queuing delays, Tor multiplexing, congestion, and padding-based defenses. 
As a result, passive correlation can become less reliable in noisy or defended settings.

Active bandwidth perturbation offers a different approach. 
Instead of relying only on naturally occurring metadata patterns, an infrastructure-level adversary can impose controlled throughput variations on selected Tor traffic. 
These low-frequency throughput patterns are less sensitive to fine-grained packet-timing jitter than packet-level timing features, although they can still be distorted by congestion, flow control, and path variability. 
The correlation task is therefore transformed from comparing noisy natural traces to detecting whether a known modulation pattern is present in the observed exit-side traffic.

Recovering such patterns remains non-trivial. 
Tor traffic is affected by onion routing, multiplexing, TCP congestion control, SENDME-based flow control, relay load, and cross-region network variability. 
These factors may smooth, shift, or partially erase the injected bandwidth pattern before it reaches the exit side. 
This motivates the use of a learning-based detector that can separate deterministic modulation patterns from natural congestion and background network variation.

\subsection{Research Questions}
\label{subsec:research_questions}

Based on this threat model, we study two research questions:

\begin{itemize}
    \item \textbf{RQ1: Feasibility and robustness.} 
    Can actively perturbed Tor flows be distinguished from natural flows under realistic network noise and congestion?

    \item \textbf{RQ2: Fine-grained correlation.} 
    To what extent can exit-side observations be associated with specific client-side modulation patterns by identifying the corresponding bandwidth-modulation geometry?
\end{itemize}

To answer these questions, we build an end-to-end analysis pipeline with four stages:

\begin{enumerate}
    \item \textbf{Signal injection.} 
    At the upstream gateway, the shaper applies one of several predefined bandwidth-modulation patterns to Tor-related traffic. 
    This produces a watermarked client-side traffic stream.

    \item \textbf{Network transmission.} 
    The watermarked traffic traverses the Tor network. 
    During transmission, the injected pattern may be distorted by onion routing, TCP dynamics, relay congestion, re-packetization, and network jitter.

    \item \textbf{Trace representation.} 
    At the exit side, observed traffic is segmented into flows and converted into fixed-length representations. 
    These representations preserve packet-level structure and temporal behavior needed for modulation detection.

    \item \textbf{Hierarchical pattern recognition.} 
The resulting representations are processed by BM-Net. 
The model first learns reusable traffic representations through self-supervised masked pre-training on serialized traffic traces, and then adapts these representations through supervised fine-tuning for binary perturbation detection and fine-grained modulation classification. 
This separation between representation learning and task-specific classification enables data-efficient adaptation when fine-grained labeled modulation traces are scarce. 
A successful classification provides evidence that the observed exit-side flow carries the modulation pattern injected at the client side.

\end{enumerate}

\section{System Architecture and Mathematical Formalism}
\label{sec:system_architecture}

This section presents the architecture and mathematical formulation of Non-invasive Active Traffic-correlation Analysis (NATA). 
NATA combines controlled bandwidth perturbation at an upstream network vantage point with exit-side pattern recognition. 
Rather than relying only on naturally occurring timing or volume similarities, NATA imposes low-frequency throughput patterns on selected Tor-related traffic and attempts to recover these patterns from noisy exit-side observations.

The main technical challenge is that the injected pattern may be distorted by Tor multiplexing, relay congestion, TCP dynamics, SENDME-based flow control, and cross-region network jitter. 
Accordingly, the system is designed around four stages: 
(1) constraint-aware traffic shaping, 
(2) hierarchical serialized representation, 
(3) selective state-space neural encoding, and 
(4) hierarchical training and inference.

\begin{figure*}[h]
    \centering
    \includegraphics[width=0.88\linewidth]{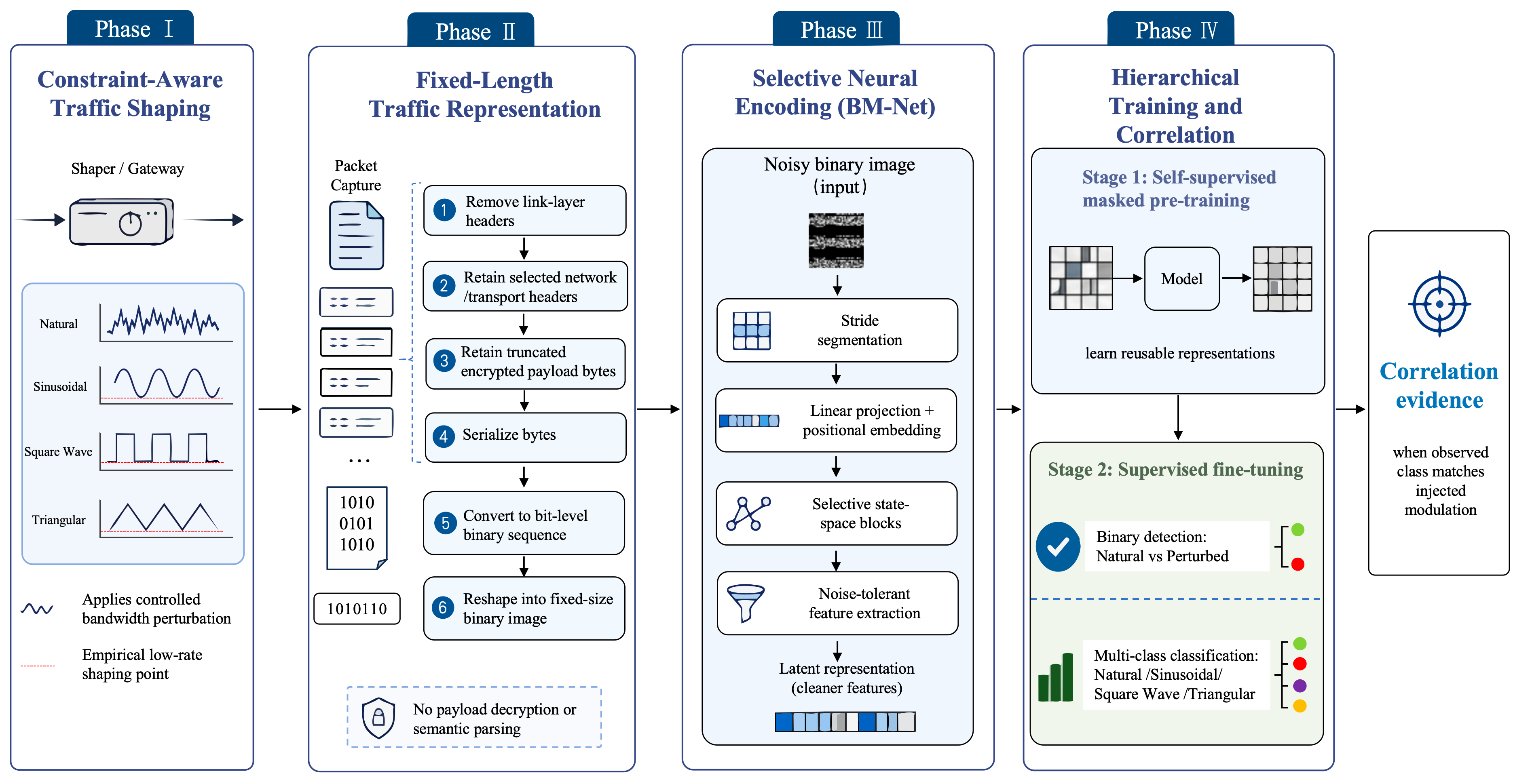}
    \caption{Overview of the NATA pipeline. The system operates in four phases: (I) active bandwidth shaping near the client-side gateway, (II) fixed-length representation of captured exit-side traffic, (III) selective neural encoding for noise-tolerant feature extraction, and (IV) hierarchical training and modulation-based traffic correlation.}
    \label{fig:system_architecture}
\end{figure*}

\subsection{Phase I: Constraint-Aware Traffic Shaping}

NATA introduces controlled bandwidth perturbations into selected Tor-related traffic at an upstream network vantage point. 
The adversary-controlled shaper applies a time-varying rate limit to the target Tor TCP connection, while avoiding endpoint compromise, Tor-browser modification, or packet-payload modification. 
To distinguish injected patterns from natural congestion, we use a small modulation dictionary consisting of deterministic low-frequency throughput patterns. 
Natural congestion may produce transient throughput drops, but it is less likely to follow a stable geometric waveform over multiple modulation periods.

\subsubsection{Modulation Dictionary and Function Definitions}

We implement the shaping mechanism using a programmable token-bucket shaper. 
Let $r(t)$ denote the configured token refill rate at time $t$. 
We define a modulation dictionary 
$\mathcal{M}=\{m_0,m_1,m_2,m_3\}$, where $m_0$ denotes natural traffic without explicit shaping, and $m_1$, $m_2$, and $m_3$ denote sinusoidal, square-wave, and triangular modulation, respectively. 
Equivalently, the multi-class setting has $K=3$ active modulation types, with class indices ranging from $0$ to $K$.

For the active classes, the target shaping rates are defined as:
\begin{align}
f_{\mathrm{sine}}(t)
&= r_{\mathrm{base}}+\mathcal{A}\sin(2\pi f_{\mathrm{mod}}t+\phi), \label{eq:f_sine}\\
f_{\mathrm{sq}}(t)
&=
\begin{cases}
r_{\mathrm{high}}, & \sin(2\pi f_{\mathrm{mod}}t)\geq 0,\\
r_{\mathrm{low}}, & \sin(2\pi f_{\mathrm{mod}}t)<0,
\end{cases}
\label{eq:f_square}\\
f_{\mathrm{tri}}(t)
&= r_{\mathrm{base}}+
\frac{2\mathcal{A}}{\pi}
\arcsin\!\left(\sin(2\pi f_{\mathrm{mod}}t+\phi)\right).
\label{eq:f_tri}
\end{align}

Here, $r_{\mathrm{base}}$ is the baseline shaping rate, $\mathcal{A}$ is the modulation amplitude, $f_{\mathrm{mod}}$ is the modulation frequency, and $\phi$ is the phase offset. 
The square-wave pattern alternates between high-rate and low-rate states, while the triangular pattern provides approximately linear ramp-up and ramp-down regions. 
The sinusoidal and triangular functions have nominal ranges $[r_{\mathrm{base}}-\mathcal{A}, r_{\mathrm{base}}+\mathcal{A}]$; in implementation, the final shaping rate is clipped to remain within the configured rate bounds.

\subsubsection{Flow-Control-Aware Rate Bounding}

Tor uses window-based flow control, and overly aggressive throttling can cause stalls, timeouts, or circuit replacement. 
Therefore, the final shaping function is bounded below by a minimum configured rate:
\begin{equation}
    r(t) = \max\left(f_{\mathrm{target}}(t), r_{\min}\right),
\end{equation}
where $f_{\mathrm{target}}(t)$ denotes the selected modulation function.

The value of $r_{\min}$ should not be interpreted as a universal Tor keep-alive threshold. 
In our experiments, the low-rate phase is selected empirically to create a visible throughput perturbation while avoiding immediate circuit failure in the collected traces.
The usable lower bound may vary with relay load, path length, TCP congestion-control behavior, Tor multiplexing, and background network conditions.

At the transport layer, rate limiting may introduce pacing, queueing delay, and, when buffers overflow, packet loss. 
These effects can interact with TCP congestion control and Tor's SENDME-based flow control. 
For this reason, the modulation period is chosen at a macroscopic timescale, so that the intended throughput pattern is less sensitive to short-term packet-level variation.

\subsection{Phase II: Hierarchical Serialized Representation}
\label{subsec:feature_representation}

NATA represents each captured flow as a fixed-length packet sequence derived from raw packet captures. 
The goal is to preserve packet-level structure while avoiding reliance on decrypted payload content or application-layer semantics. 
Although the final byte sequence is converted into a two-dimensional binary image for neural processing, the construction preserves the original packet and byte order before tokenization.

\subsubsection{Flow Serialization and Normalization}

Each bidirectional flow is extracted from raw packet captures and represented as an ordered sequence of packets. 
For each packet, we remove data-link-layer headers to reduce device-specific artifacts, while retaining selected network- and transport-layer header bytes. 
We also retain a truncated segment of encrypted payload bytes. 
The retained payload bytes are used only as raw encrypted byte patterns: the model does not decrypt payloads, parse application-layer semantics, or inspect user content.

To obtain a fixed-length representation, we retain the first $M$ packets of each flow. 
For each retained packet, the header bytes and encrypted-payload bytes are separately truncated or zero-padded to predefined lengths. 
Flows with fewer than $M$ packets are padded at the packet level, whereas flows with more than $M$ packets are truncated to the first $M$ packets.

The resulting byte sequence is expanded into a bit-level binary sequence by representing each byte with its 8-bit encoding. 
For neural processing, this binary sequence is arranged into a fixed-size two-dimensional binary representation. 
The subsequent tokenization follows the original serialized order, preserving packet order and local byte-level structure.

In addition to the bit-level representation, we record packet timestamps, sizes, and directions to compute temporal statistics such as inter-arrival times and per-bin throughput. 
These statistics are derived from the same packet captures and are used for trace characterization and visualization.

\subsubsection{Context-Aware Stride Segmentation}

Directly processing the full binary representation can be computationally expensive. 
We therefore divide the serialized representation into non-overlapping strides of length $L_s$. 
If the sequence length is not divisible by $L_s$, zero padding is applied to the final stride:
\begin{equation}
    \mathcal{S} = \{ \mathbf{s}_1, \mathbf{s}_2, \dots, \mathbf{s}_N \},
    \qquad
    \mathbf{s}_i \in \{0,1\}^{L_s}.
\end{equation}
This stride-based tokenization reduces the effective sequence length while preserving local byte- and packet-order information.

\subsubsection{Linear Projection and Positional Encoding}

Each stride $\mathbf{s}_i$ is projected into a $D_{\mathrm{model}}$-dimensional latent space using a learnable linear projection:
\begin{equation}
    \mathbf{x}_i^{(0)} = \mathbf{s}_i \mathbf{W}_{\mathrm{proj}} + \mathbf{p}_i,
\end{equation}
where $\mathbf{W}_{\mathrm{proj}} \in \mathbb{R}^{L_s \times D_{\mathrm{model}}}$ and $\mathbf{p}_i$ is a learnable positional embedding. 
The positional embedding preserves the ordering of stride tokens in the serialized flow representation.

\subsection{Phase III: Selective State-Space Neural Encoder}

The main challenge in exit-side detection is that the injected throughput pattern is distorted by network jitter, relay congestion, Tor multiplexing, and transport-layer dynamics. 
Simple correlation metrics may be sensitive to local timing shifts and packet-level burstiness. 
BM-Net therefore builds on a selective state-space traffic encoder to model long-range sequential dependencies with linear-time recurrence, and adapts this encoder to the bandwidth-modulation detection setting.

The motivation for using a selective state-space model is that bandwidth modulation is a low-frequency structure, whereas much of the packet-level variation is short-term noise. 
The selective recurrence allows the model to condition its state updates on the current input token, which helps it emphasize tokens that are informative for the modulation pattern and downweight less informative local fluctuations.

\subsubsection{Continuous-Time System Abstraction}

We model the latent state of the traffic representation as a continuous-time linear system:
\begin{equation}
    h'(t) = \mathbf{A}h(t) + \mathbf{B}x(t), 
    \qquad 
    y(t) = \mathbf{C}h(t),
\end{equation}
where $h(t)$ is the latent state, $x(t)$ is the input, $y(t)$ is the output representation, and $\mathbf{A}$ governs the state transition dynamics. 
This formulation provides a structured way to summarize long-range flow history in a finite-dimensional latent state.

\subsubsection{Discretization via Zero-Order Hold}

To process discrete packet-derived tokens, the continuous-time system is discretized using a token-dependent timescale parameter $\Delta_t$. 
Under zero-order hold discretization, the transition parameters for token $t$ are:
\begin{equation}
\overline{\mathbf{A}}_t = \exp(\Delta_t \mathbf{A}), 
\qquad
\overline{\mathbf{B}}_t =
\left(
\int_{0}^{\Delta_t}
\exp(\tau \mathbf{A})\,d\tau
\right)
\mathbf{B}_t .
\end{equation}
When $\mathbf{A}$ is invertible, the integral form is equivalent to 
$\mathbf{A}^{-1}(\exp(\Delta_t\mathbf{A})-\mathbf{I})\mathbf{B}_t$. 
We use the integral notation to avoid imposing an unnecessary invertibility assumption on $\mathbf{A}$.

The resulting discrete recurrence is:
\begin{equation}
\mathbf{h}_t =
\overline{\mathbf{A}}_t\mathbf{h}_{t-1}
+
\overline{\mathbf{B}}_t\mathbf{x}_t,
\qquad
\mathbf{y}_t = \mathbf{C}_t\mathbf{h}_t .
\end{equation}
This recurrence summarizes the serialized flow representation in a finite-dimensional latent state while allowing the update dynamics to vary across input tokens.

\subsubsection{Selective Scan Mechanism}

Standard state-space models use fixed transition parameters across tokens. 
However, network traffic is non-stationary: bursts, idle periods, congestion events, and shaped intervals may have different relevance to modulation detection. 
BM-Net therefore uses input-dependent parameters:
\begin{equation}
\begin{aligned}
    \mathbf{B}_t &= \mathrm{Linear}_B(\mathbf{x}_t), \\
    \mathbf{C}_t &= \mathrm{Linear}_C(\mathbf{x}_t), \\
    \Delta_t &= \mathrm{softplus}
    \left(\mathrm{Linear}_{\Delta}(\mathbf{x}_t)\right).
\end{aligned}
\end{equation}
The softplus function keeps the timescale parameter positive. 
The resulting selective scan updates the latent state according to the content of each token. 
This allows the model to emphasize information relevant to macroscopic modulation patterns while reducing sensitivity to short-term packet-level variation.

\subsection{Phase IV: Hierarchical Training and Multi-Class Inference}

High-fidelity labeled Tor traces are difficult to collect at scale, especially across long-haul paths and different modulation settings. 
We therefore use a two-stage training strategy that separates representation learning from task-specific classification.

\subsubsection{Stage I: Self-Supervised Masked Pre-Training}

The backbone is first trained using a self-supervised masked pre-training objective on serialized traffic representations. 
During pre-training, a large fraction of input tokens is masked, and the model learns to infer useful structural information from the remaining context. 
This stage encourages the encoder to capture packet-order structure, byte-level regularities, and long-range dependencies in encrypted traffic without requiring modulation labels.

\subsubsection{Stage II: Supervised Fine-Tuning}

The pre-trained backbone is then transferred to the target classification task. 
A task-specific classification head outputs a logit vector $\mathbf{z} \in \mathbb{R}^{K+1}$, where the class set includes natural traffic and $K$ active modulation types. 
The class probability is computed using the softmax function:
\begin{equation}
    P(y=k \mid \mathbf{X}) =
    \frac{\exp(z_k)}{\sum_{j=0}^{K} \exp(z_j)}.
\end{equation}
The model is optimized using the categorical cross-entropy loss:
\begin{equation}
    \mathcal{L} =
    - \sum_{k=0}^{K}
    \mathbb{I}(y=k)
    \log P(y=k \mid \mathbf{X}).
\end{equation}

This transfer strategy reduces the amount of task-specific labeled data required for modulation identification. 
Rather than learning all representations from the smaller fine-grained dataset, the model reuses structural representations learned during masked pre-training and adapts them to distinguish among natural, sinusoidal, square-wave, and triangular modulation classes.

\section{Theoretical Modeling of Bandwidth-Perturbation Analysis}
\label{sec:theoretical_modeling}

In this section, we provide a probabilistic model for estimating the traffic-correlation success probability of NATA. 
The model separates three factors: the probability that a watermarked flow is observed by an adversary-controlled exit relay, the probability that the perturbation is detected, and the probability that the modulation type is correctly classified. 
The resulting formulation is intended to provide a first-order estimate of correlation risk under the threat model defined in Section~\ref{sec:threat}.

\subsection{Problem Formalization and Notation}

We consider an infrastructure-level adversary that can apply bandwidth perturbation to Tor-related traffic near the client side and can observe traffic at a set of adversary-controlled exit relays. 
The adversary does not compromise endpoints, decrypt payloads, inject application-layer content, or modify packet payloads. 
Bandwidth shaping may indirectly affect packet timing, pacing, retransmissions, or queueing behavior, but the adversary's direct control is limited to rate modulation.

The analysis pipeline consists of two sequential inference stages:

\begin{itemize}
    \item \textbf{Stage I: Perturbation detection.} 
    Given an observed flow, determine whether it has been subject to adversarial bandwidth perturbation. 
    Let $p_1$ denote the probability that a perturbed flow is correctly detected.

    \item \textbf{Stage II: Perturbation classification.} 
    Conditioned on successful detection, classify the flow into one of $K$ predefined bandwidth-perturbation types. 
    Let $p_{2,i}$ denote the conditional probability of correctly classifying perturbation type $i$, where $i \in \{1,\dots,K\}$.
\end{itemize}

Here, $K$ denotes the number of active perturbation types and does not include the natural-traffic class.

\subsection{Exit-Relay Observation Probability}

Tor selects exit relays using bandwidth-weighted path selection subject to relay flags, exit policies, and other path-selection constraints. 
As a first-order approximation, we model the probability that a circuit uses an adversary-controlled exit relay according to the fraction of exit bandwidth controlled by the adversary.

Let $\mathcal{E}_{\mathrm{all}}$ denote the set of available exit relays, and let $\mathcal{E}_{\mathrm{bad}}(n) \subseteq \mathcal{E}_{\mathrm{all}}$ denote the subset of $n$ adversary-controlled exit relays. 
Let $\mathrm{BW}_j$ be the consensus bandwidth weight of exit relay $j$. 
The exit observation probability is then approximated as:
\begin{equation}
p_{\mathrm{exit}}(n)
=
\frac{
    \sum\limits_{j \in \mathcal{E}_{\mathrm{bad}}(n)}
    \mathrm{BW}_j
}{
    \sum\limits_{k \in \mathcal{E}_{\mathrm{all}}}
    \mathrm{BW}_k
}.
\label{eq:p_exit}
\end{equation}

This probability depends on both the number and bandwidth weights of adversary-controlled exits, and is instantiated using \textit{tornettools}-based simulations in Section~\ref{sec:network_sim}.

\subsection{Per-Client Traffic-Correlation Probability}

For a flow using perturbation type $i$, a successful correlation requires three events: the flow must traverse an adversary-controlled exit relay, the perturbation must be detected, and the perturbation type must be correctly classified. 
Under the simplifying assumption that these components are independent, the per-flow success probability is:
\begin{equation}
q_i(n)
=
p_{\mathrm{exit}}(n) \, p_1 \, p_{2,i}.
\label{eq:q_i}
\end{equation}

Suppose the adversary monitors $r$ independent candidate client-side flows using the same perturbation type $i$. 
The probability that at least one of these flows is successfully correlated is:
\begin{equation}
P_{\mathrm{corr}}^{(i)}(r,n)
=
1 -
\left(
1 - q_i(n)
\right)^{r}.
\label{eq:p_corr_single}
\end{equation}

This expression captures an aggregation effect: increasing the number of monitored flows or the adversarial exit bandwidth increases the overall success probability. 
However, the marginal gain decreases as $r$ grows, since $1-(1-q)^r$ is an increasing concave function of $r$ for $0<q<1$.

\subsection{Mixed Perturbation Strategies}

In a mixed strategy, different flows may use different perturbation types. 
If the adversary observes $r_i$ candidate flows using perturbation type $i$, with $\sum_{i=1}^{K} r_i = r$, the overall success probability becomes:
\begin{equation}
P_{\mathrm{corr}}
=
1 -
\prod_{i=1}^{K}
\left(
1 - q_i(n)
\right)^{r_i}.
\label{eq:p_corr_mixed_count}
\end{equation}

If perturbation types are assigned independently according to a distribution $\{\pi_1,\dots,\pi_K\}$, where $\sum_{i=1}^{K}\pi_i=1$, the expected per-flow success probability is:
\begin{equation}
q_{\mathrm{mix}}(n)
=
p_{\mathrm{exit}}(n) \, p_1
\sum_{i=1}^{K}
\pi_i p_{2,i}.
\label{eq:q_mix}
\end{equation}
The corresponding success probability over $r$ independently monitored flows is:
\begin{equation}
P_{\mathrm{corr,mix}}(r,n)
=
1 -
\left(
1 - q_{\mathrm{mix}}(n)
\right)^r.
\label{eq:p_corr_mixed}
\end{equation}

This formulation captures heterogeneous classification difficulty across perturbation types through the class-specific term $p_{2,i}$.

\subsection{Temporal Aggregation}

Finally, we consider repeated observation over multiple time windows. 
Let $P_t$ denote the traffic-correlation success probability in observation window $t$. 
Under a simplifying independence assumption across observation windows, the cumulative success probability over $T$ windows is:
\begin{equation}
P_{\mathrm{corr,total}}
=
1 -
\prod_{t=1}^{T}
\left(1-P_t\right).
\label{eq:p_corr_temporal_general}
\end{equation}

If each window has the same success probability $P_{\mathrm{corr}}$, this reduces to:
\begin{equation}
P_{\mathrm{corr,total}}
=
1 -
\left(
1 - P_{\mathrm{corr}}
\right)^T.
\label{eq:p_corr_temporal}
\end{equation}

This temporal aggregation model shows that repeated observations can increase correlation probability even when the single-window success probability is modest. 
The independence assumption is an idealization: in practice, successive windows may share the same circuit, relay path, congestion state, or classification errors. 
Therefore, we interpret Eq.~\ref{eq:p_corr_temporal} as a first-order approximation rather than a guarantee of independent evidence accumulation.

\section{System Analysis and Evaluation}
\label{sec:evaluation}

In this section, we empirically instantiate the probability model introduced in Section~\ref{sec:theoretical_modeling}. 
Specifically, we estimate the network-layer exit-observation component $p_{\mathrm{exit}}$ using \textit{tornettools}-based Tor simulations, and measure the detection and classification components $p_1$ and $p_{2,i}$ using real-world Tor traffic traces. 
Together, these components allow us to assess traffic-correlation risk under the considered infrastructure-level adversary model.

The evaluation proceeds as follows. 
Section~\ref{sec:network_sim} presents the network simulation and exit-observation analysis. 
Section~\ref{sec:setup} describes the real-world experimental setup and data collection. 
Section~\ref{sec:implementation} details the BM-Net implementation and training procedure. 
Section~\ref{sec:results} reports the classification results and baseline comparisons. 
Section~\ref{sec:sensitivity} analyzes robustness under challenging network conditions and client-side defenses.

\subsection{Network Simulation and Exit-Observation Analysis}
\label{sec:network_sim}

To estimate the network-layer component $p_{\mathrm{exit}}(n)$ in Eq.~\ref{eq:p_exit}, we conducted a scaled-down simulation of the Tor network using tornettools. 
The goal of this simulation is to approximate the probability that a watermarked Tor connection is routed through an adversary-controlled exit relay under Tor's bandwidth-weighted path-selection mechanism.

\subsubsection{Simulation Setup with Tornettools}

Directly simulating the full Tor network is computationally expensive. 
We therefore used a 1\% scaled tornettools network constructed from Tor consensus data. 
The scaled network preserves the bandwidth-weighted relay selection mechanism and approximates the relative bandwidth distribution used in Tor path selection. 
This makes it useful for estimating first-order exit-observation probability.

However, the scaling process also reduces the absolute number of relays and circuits, which may affect path diversity, relay-collision probability, and the tail behavior of high-bandwidth exits. 
Therefore, we interpret the simulated exit-observation probability as an approximation of the full-network trend rather than an exact reproduction of the live Tor network.

\begin{table}[h]
\centering
\caption{Comparison Between Real and Simulated Tor Networks}
\label{tab:sim-overview}
\begin{tabular}{lcc}
\toprule
\textbf{Category} & \textbf{Real Network} & \textbf{Simulated Network} \\
\midrule
Number of relays & 7,884 & 80 \\
Total bandwidth & 1053.5 Gbit/s & Scaled and normalized \\
User agents & Unknown & 11,121 \\
Scaling factor & Full scale & 0.01 \\
\bottomrule
\end{tabular}
\end{table}

The simulated network consists of 80 relays, including 14 Exit-Guard relays and 9 Exit relays, and generates approximately 14,900 circuits every 10 minutes.

\subsubsection{Malicious Exit Injection Results}

We progressively replaced benign exit relays with adversary-controlled relays whose bandwidth ranged from 27 Mbps to 148 Mbps. 
We define the empirical exit-observation probability $\hat{p}_{\mathrm{exit}}$ as the fraction of simulated circuits for which the final hop is an adversary-controlled exit relay.

As the number of adversary-controlled exits increased from 1 to 9, the empirical exit-observation probability increased monotonically:
\begin{itemize}
    \item With 1 adversary-controlled exit relay of 148 Mbps, $\hat{p}_{\mathrm{exit}} \approx 2.13\%$.
    \item With 5 adversary-controlled exit relays, $\hat{p}_{\mathrm{exit}}$ exceeded 10\%.
\end{itemize}

These empirical results are consistent with the bandwidth-weighted trend predicted by Eq.~\ref{eq:p_exit}, suggesting that higher-bandwidth adversary-controlled exits are more likely to attract Tor circuits under the simulated setting.

\subsection{Real-World Experimental Setup and Data Collection}
\label{sec:setup}

\subsubsection{Bandwidth Perturbation Mechanism}

To implement non-invasive active throttling, we use \texttt{wondershaper}, a lightweight traffic-control utility built on the Linux \texttt{tc} subsystem. 
Rather than relying on deep packet inspection, the shaper enforces interface-level rate limits through standard queuing disciplines. 
We deployed adversary-controlled Tor exit relays on high-bandwidth AWS instances and designed three distinct modulation fingerprints to evaluate multi-flow correlation.

\begin{enumerate}
    \item \textbf{Square-wave modulation.}
    This pattern approximates a binary on-off keying signal. 
    It alternates between an unrestricted baseline state, where throughput is governed by the natural capacity of the Tor circuit, and a low-rate state. 
    In our implementation, the modulation period is set to 30 seconds.

    \item \textbf{Sinusoidal modulation.}
    This pattern introduces smooth periodic rate variation, allowing us to evaluate whether the model can recover continuous low-frequency throughput changes after Tor-induced distortion.

    \item \textbf{Triangular modulation.}
    This pattern serves as a stress-test case. 
    Unlike the smooth sinusoidal waveform, the triangular waveform contains approximately linear rising and falling slopes as well as sharper turning points.
\end{enumerate}

\begin{figure}[h]
    \centering
    \includegraphics[width=1.0\linewidth]{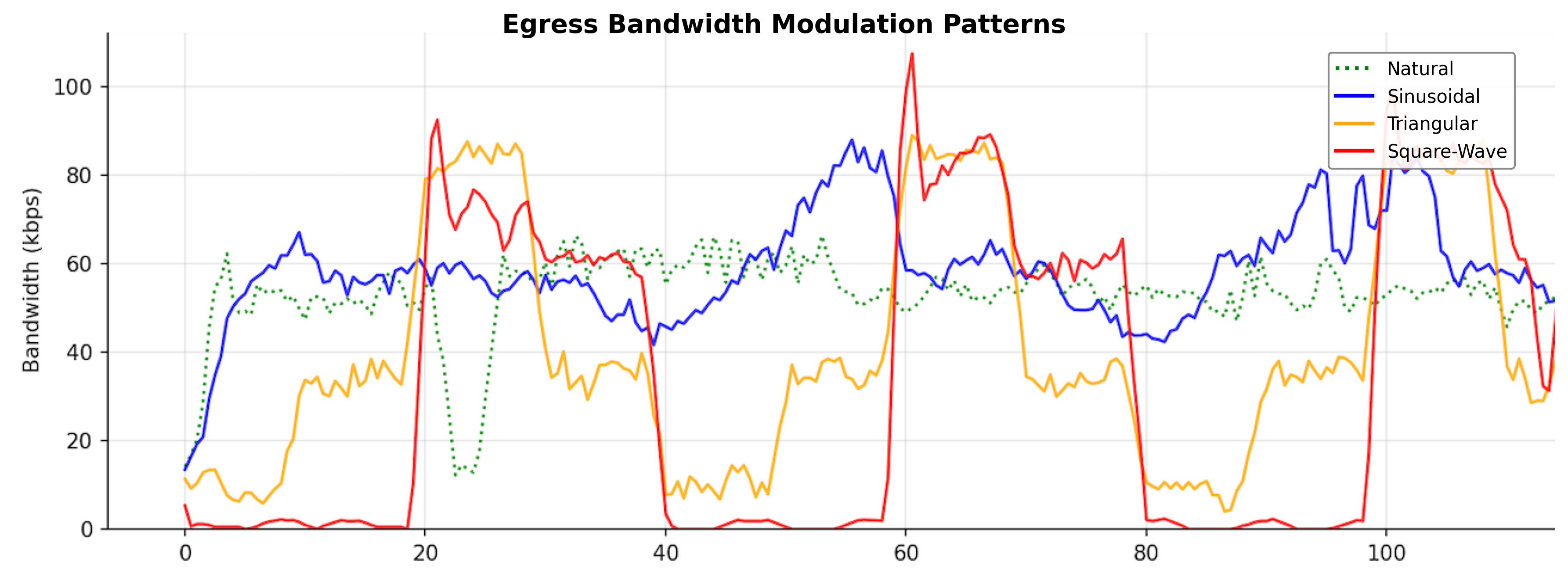} 
    \caption{Egress bandwidth modulation patterns. The plot illustrates the captured bandwidth dynamics for natural, square-wave, sinusoidal, and triangular traffic. The watermarked flows exhibit distinguishable geometric patterns compared with natural traffic.}
    \label{fig:modulation_patterns}
\end{figure}

Figure~\ref{fig:modulation_patterns} visualizes the throughput signatures of these modulation schemes as observed at the exit relay. 
Despite network jitter, the square wave retains abrupt transition regions, while the triangular wave preserves approximately linear rising and falling trends.

\subsubsection{Client Deployment and Traffic Generation}

To cover diverse network paths, we deployed ten Tor client instances across five geographic regions: Europe, North America, Australia, Southeast Asia, and East Asia. 
We used a large website list as the target domain pool and retained 100,000 reachable websites after accessibility filtering. 
Traffic was generated using \texttt{tor-selenium}, which automates full-page loading through Tor.

\subsubsection{Dataset Construction}

The data-collection campaign lasted two weeks. 
Raw packet captures were preprocessed into fixed-length input representations. 
For each flow, we extract packet headers and truncate a fixed number of bytes from each packet payload, following the representation design in Section~\ref{sec:system_architecture}. 
The resulting byte sequence is converted into a binary image representation and used as the model input. 
We emphasize that the payload bytes are treated only as raw encrypted byte patterns: the classifier does not decrypt payloads, parse application-layer semantics, or inspect user content.

\begin{figure}[h]
    \centering
    \includegraphics[width=1.0\linewidth]{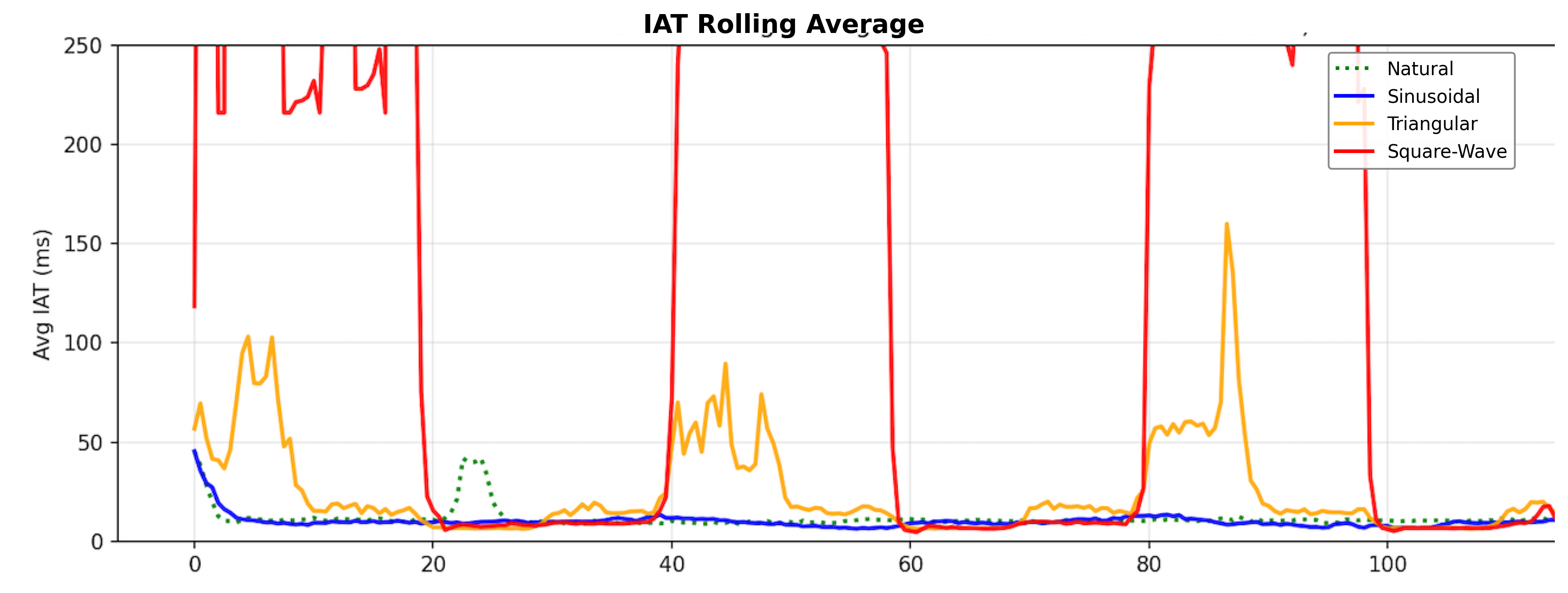} 
    \caption{Feature analysis of the rolling average of inter-arrival times. The plot shows temporal patterns associated with natural, square-wave, sinusoidal, and triangular traffic, providing intuition about how bandwidth modulation affects packet timing.}
    \label{fig:iat_feature}
\end{figure}

In addition to the byte-level representation, we compute temporal statistics such as inter-arrival times, packet sizes, and per-bin throughput from the same packet captures. 
These features are used for trace characterization and visualization, as shown in Figure~\ref{fig:iat_feature}. 
These statistics are derived only from packet timestamps and sizes rather than from decrypted content.

To evaluate robustness against client-side defenses, a subset of traces was collected under a hybrid defense configuration combining Walkie-Talkie and WTF-PAD. 
Walkie-Talkie reshapes traffic at the burst level, while WTF-PAD injects adaptive padding at the timing level. 
Including these defended traces helps evaluate whether the model learns the deterministic throughput constraints induced by bandwidth shaping rather than relying only on superficial packet-level artifacts.

The resulting dataset contains 20,000 traces balanced across four classes: natural, square-wave, sinusoidal, and triangular traffic. 
Unless otherwise specified, the dataset is split at the flow level: packets or segments from the same flow are not shared across training and testing sets. 
This reduces the risk of train-test leakage and prevents the classifier from memorizing flow-specific artifacts. 
The classification target is the modulation type rather than the visited website.

To better understand the temporal characteristics of these traces, Figure~\ref{fig:iat_feature} shows the rolling average of inter-arrival times. 
Triangular modulation produces sustained regions with relatively stable packet intervals due to gradual bandwidth changes, whereas square-wave modulation creates sharper transitions. 
Natural traffic exhibits more irregular temporal variation.

\subsection{Model Implementation and Training}
\label{sec:implementation}

BM-Net is implemented in PyTorch using a selective state-space traffic encoder adapted from a NetMamba-style backbone~\cite{wang2024netmamba}, with task-specific modifications for bandwidth-modulation detection and classification.
These modifications include the serialized binary flow representation, the bandwidth-modulation label space, and the self-supervised pre-training followed by supervised fine-tuning pipeline described in Section~\ref{sec:system_architecture}. 
All models are trained and evaluated using the same flow-level data split.

\paragraph{Dataset partitioning}
To reduce train-test leakage, we partition the dataset at the flow level rather than the packet or segment level. 
Specifically, 80\% of the captured flows are used for training and validation, and the remaining 20\% are reserved for testing. 
Packets or segments from the same flow are not shared across training and testing sets.

\paragraph{Training pipeline}
BM-Net is trained in two stages. 
The first stage performs self-supervised masked pre-training on the serialized traffic representation. 
During this stage, a large fraction of input tokens is masked, and the encoder is trained to recover useful structural representations from the remaining context. 
This objective encourages the model to learn packet-order, byte-level, and long-range dependency patterns without requiring fine-grained modulation labels.

The second stage performs supervised fine-tuning for the target classification task. 
The pre-trained encoder weights are initialized from the self-supervised checkpoint, and a task-specific classification head is attached. 
For binary detection, the head distinguishes natural traffic from perturbed traffic. 
For fine-grained modulation classification, the head predicts the corresponding traffic class among natural, square-wave, sinusoidal, and triangular traces.

\paragraph{Hyperparameters}
In the pre-training stage, we train the encoder for 150,000 steps with a batch size of 128, a base learning rate of $10^{-3}$, and a mask ratio of 0.9. 
Automatic mixed precision is disabled during training. 
In the fine-tuning stage, we train for 120 epochs with a base learning rate of $2\times 10^{-3}$ and set the number of output classes according to the target task. 
The same pre-trained checkpoint is used to initialize the fine-tuning runs for binary detection and multi-class modulation classification.

\subsubsection{Baseline Configurations}

We compare BM-Net with representative traffic-analysis models from the website-fingerprinting and flow-correlation literature. 
All baselines are trained and evaluated on the same flow-level data partitions as BM-Net. 
The baseline implementations are based on the unified WFlib framework~\cite{deng2024wflib}, which provides reproducible implementations of multiple website-fingerprinting models. 
Where possible, we follow the feature representations and training settings reported in the original papers or public implementations, and tune only basic training parameters such as learning rate, batch size, and sequence length to ensure stable convergence on our dataset.

Because many of these baselines were originally designed for passive website fingerprinting rather than active bandwidth-modulation detection, the comparison should be interpreted as a reference point rather than a direct replacement study. 
Table~\ref{tab:baseline_configurations} summarizes the input representation and truncation length used for each baseline. 
For packet-sequence baselines, the length denotes the number of packets retained after truncation; for feature-based baselines, it denotes the number of extracted feature steps used as model input.

\begin{table}[t]
\centering
\caption{Summary of baseline input configurations.}
\label{tab:baseline_configurations}
\resizebox{\columnwidth}{!}{%
\begin{tabular}{lcc}
\toprule
Model & Input / Feature Type & Input Length / Truncation \\
\midrule
ARES~\cite{shen2023subverting} & MTAF features & 8000 feature steps \\
DF~\cite{sirinam2018deep} & Direction sequence & 5000 packets \\
AWF~\cite{Rimmer2018} & Direction sequence & 3000 packets \\
BAPM~\cite{guan2021bapm} & Direction sequence & 8500 packets \\
Holmes~\cite{deng2024wflib} & Temporal aggregated features & 2000 feature steps \\
NetCLR~\cite{bahramali2023realistic} & Direction sequence & 5000 packets \\
RF~\cite{deng2023robust} & Timing metrics & 1800 feature steps \\
TF~\cite{sirinam2019triplet} & Direction sequence & 5000 packets \\
TikTok~\cite{rahman2020tik} & Dense timing features & 5000 feature steps \\
TMWF~\cite{jin2023tmwf} & Direction sequence & 30720 packets \\
VarCNN~\cite{bhat19} & Direction + timing & 5000 packets \\
\bottomrule
\end{tabular}
}
\end{table}

\subsection{Classification Results and Analysis}
\label{sec:results}

We evaluate performance using standard metrics on a held-out test set comprising the remaining 20\% of flows.

\subsubsection{Stage I: Binary Anomaly Detection}
The first objective is to distinguish between Natural Traffic and Watermarked Traffic.

\begin{table}[h]
\centering
\caption{Stage I performance of binary perturbation detection.}
\label{tab:binary_results}
\begin{tabular}{lccc}
\toprule
Class & Precision & Recall & F1 Score \\
\midrule
Natural & 99.70\% & 99.59\% & 99.64\% \\
Watermarked & 99.60\% & 99.70\% & 99.65\% \\
\midrule
Overall & 99.65\% & 99.65\% & 99.65\% \\
\bottomrule
\end{tabular}
\end{table}

As shown in Table~\ref{tab:binary_results}, BM-Net achieves an overall F1 score of 99.65\% for binary perturbation detection under our evaluated setting. 
This high score is consistent with the active nature of the attack: the bandwidth shaper introduces a deterministic throughput constraint that is more separable from natural traffic than the subtle statistical leakage used in passive traffic analysis.

\begin{figure}[h]
    \centering
    \includegraphics[width=1.0\linewidth]{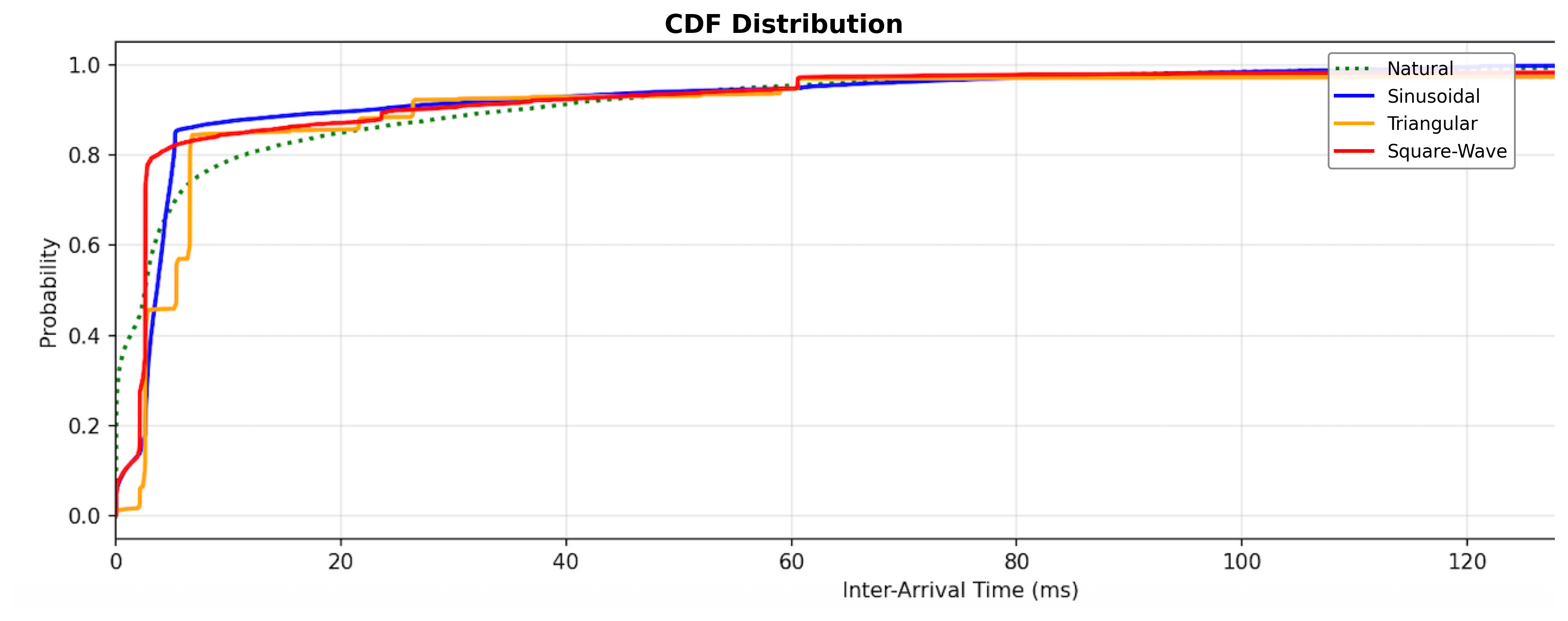} 
  \caption{Statistical feature analysis. The top pane shows raw bandwidth traces, the middle pane shows the rolling average of inter-arrival times, and the bottom pane shows the CDF of inter-arrival times. The distributions illustrate differences between natural and shaped traffic under the evaluated setting.}
    \label{fig:cdf_distribution}
\end{figure}

Figure~\ref{fig:cdf_distribution} provides additional intuition. 
The watermarked traces show a more concentrated inter-arrival-time distribution in the low-latency region, which is consistent with burst release after rate limiting. 
In contrast, natural traffic exhibits a smoother and more dispersed distribution. 
These observations suggest that the binary task is easier than fine-grained modulation classification, since it only requires distinguishing shaped traffic from natural traffic rather than separating similar waveform geometries.

For BM-Net, the binary test confusion matrix contains 982 correctly classified natural traces and 986 correctly classified watermarked traces, with only 7 errors out of 1,975 test samples. 
Because the two classes are nearly balanced and the errors are almost symmetric, the weighted-average precision, recall, and F1 score round to the same value of 99.65\%.

These results should be interpreted in light of the active nature of the setting. 
The binary task distinguishes natural traffic from traffic affected by an imposed throughput constraint, whereas passive website fingerprinting typically relies on weaker naturally occurring metadata patterns. 
Therefore, the high binary detection score does not imply that fine-grained modulation classification or passive traffic analysis is equally easy.

\paragraph{Baseline Comparison and Architectural Justification}

\begin{table}[h]
\centering
\caption{Binary perturbation-detection performance compared with baseline models. For BM-Net, precision, recall, and F1 are weighted-average metrics.}
\label{tab:baseline_comparison}
\resizebox{\columnwidth}{!}{%
\begin{tabular}{lcccc}
\toprule
Model & Accuracy & Precision & Recall & F1 Score \\
\midrule
TMWF & 50.25\% & 25.13\% & 50.00\% & 33.44\% \\
ARES & 52.68\% & 52.68\% & 52.68\% & 52.67\% \\
Holmes & 53.69\% & 53.77\% & 53.71\% & 53.54\% \\
AWF & 55.70\% & 55.70\% & 55.68\% & 55.66\% \\
RF & 58.04\% & 58.11\% & 58.06\% & 57.98\% \\
BAPM & 63.65\% & 64.03\% & 63.61\% & 63.36\% \\
NetCLR & 64.24\% & 64.24\% & 64.23\% & 64.23\% \\
TF & 64.82\% & 64.82\% & 64.82\% & 64.82\% \\
DF & 64.99\% & 65.24\% & 65.02\% & 64.88\% \\
VarCNN & 74.46\% & 74.47\% & 74.45\% & 74.45\% \\
TikTok & 75.96\% & 75.96\% & 75.96\% & 75.96\% \\
\midrule
BM-Net (Ours) & 99.65\% & 99.65\% & 99.65\% & 99.65\% \\
\bottomrule
\end{tabular}
}
\end{table}

As shown in Table~\ref{tab:baseline_comparison}, BM-Net outperforms the evaluated baseline models on the binary perturbation-detection task. 
This gap is partly due to the mismatch between the design assumptions of the baselines and the target task considered in this paper. 
Many baseline models were developed for passive website fingerprinting, where the discriminative information is often encoded in packet direction, burst shape, or local timing patterns. 
By contrast, active bandwidth-modulation detection depends on recovering low-frequency throughput structures that may be distorted by Tor multiplexing, congestion, long-haul jitter, and client-side traffic-analysis defenses.

The defense-aware nature of our dataset further increases the difficulty of this task. 
The training and testing traces include traffic collected under WTF-PAD and Walkie-Talkie, which modify timing patterns, padding behavior, and burst-level structure. 
These transformations can weaken the local packet-level cues used by many passive-analysis baselines. The flow-level split preserves this mixed distribution, so both training and testing contain undefended and defended traces.
As a result, models that rely primarily on packet directions, burst shapes, or short-range timing features may fail to recover the more macroscopic throughput constraints introduced by active bandwidth perturbation.

Another important factor is data efficiency. 
Most baselines are trained end-to-end on the available labeled target data, which makes them sensitive to small-sample regimes and distribution shifts. 
BM-Net instead uses self-supervised masked pre-training to learn reusable sequential representations before supervised fine-tuning. 
This design is particularly useful when fine-grained labeled modulation traces are scarce and when the dataset contains both natural Tor variability and defense-induced traffic transformations, as in our multi-class evaluation.

\subsubsection{Stage II: Fine-Grained Modulation Identification}

We next evaluate whether BM-Net can distinguish among specific modulation geometries. 
This task is more challenging than binary detection because the model must separate natural, square-wave, sinusoidal, and triangular traffic under real-world network distortion.

\paragraph{Transfer learning strategy}
The fine-grained dataset is smaller than the binary detection dataset because collecting high-quality labeled traces across cross-region Tor paths is costly and unstable. 
We therefore initialize the encoder using the weights learned during self-supervised masked pre-training and fine-tune the model on the four-class modulation task.

\paragraph{Identification results}
As shown in Table~\ref{tab:multiclass_results}, BM-Net achieves a macro-F1 score of 97.5\% on the fine-grained modulation classification task. 
The strongest performance is observed for sinusoidal and triangular traffic, while most remaining errors occur between natural and square-wave traffic. 
This suggests that abrupt square-wave transitions may be partially smoothed by network dynamics, making them more difficult to distinguish under some paths.

\begin{table}[h]
\centering
\caption{Fine-grained modulation identification performance.}
\label{tab:multiclass_results}
\resizebox{\columnwidth}{!}{%
\begin{tabular}{lccc}
\toprule
Modulation Type & Precision & Recall & F1 Score \\
\midrule
Natural Traffic & 94.5\% & 98.1\% & 96.3\% \\
Sinusoidal      & 98.1\% & 100.0\% & 99.0\% \\
Square Wave     & 97.8\% & 93.6\% & 95.7\% \\
Triangular      & 100.0\% & 98.0\% & 99.0\% \\
\midrule
Macro Average (BM-Net) & 97.6\% & 97.4\% & 97.5\% \\
\bottomrule
\end{tabular}
}
\end{table}

\begin{figure}[h]
    \centering
    \includegraphics[width=1.0\linewidth]{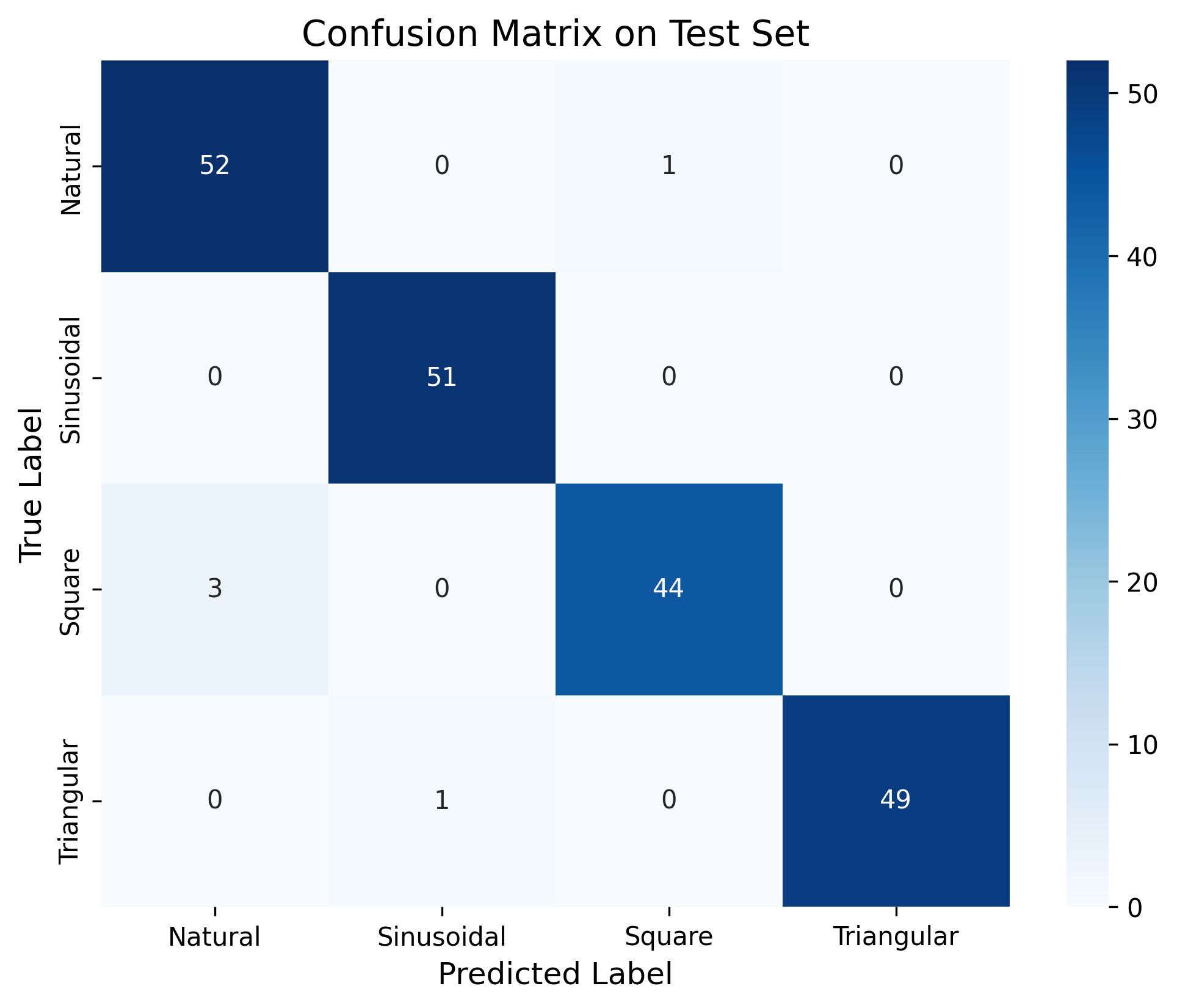}
    \caption{Confusion matrix counts for modulation classification on the 201-sample held-out test set. The model separates sinusoidal and triangular traffic with low confusion, while most remaining errors occur between natural and square-wave traffic.}
    \label{fig:confusion_matrix}
\end{figure}

\paragraph{Confusion-matrix analysis}
Figure~\ref{fig:confusion_matrix} shows the confusion matrix counts for the four-class task.
The model separates sinusoidal and triangular traffic with low confusion, while the main errors occur between natural and square-wave traffic. 
This pattern is consistent with the intuition that abrupt square-wave transitions may be smoothed by network dynamics, making weak or distorted square-wave traces harder to separate from natural congestion.

\subsubsection{Small-Sample Adaptation under Dataset Shift}

A central advantage of BM-Net is its ability to adapt to fine-grained modulation classification with limited labeled data. 
Unlike many baseline models that are trained end-to-end on the target dataset, BM-Net separates representation learning from task-specific classification. 
The self-supervised masked pre-training stage is performed on a larger traffic dataset without using fine-grained modulation labels, while the supervised fine-tuning stage uses a much smaller labeled dataset for the four-class modulation task.

This setting introduces both data scarcity and dataset shift: the pre-training and fine-tuning data are collected under different conditions and serve different learning objectives. 
The fine-grained test set used for the confusion matrix contains 201 held-out samples, reflecting the practical difficulty of collecting reliable multi-class ground truth over live Tor paths. 
Despite this limited supervision, BM-Net achieves a 97.5\% macro-F1 score, suggesting that the masked pre-training stage learns reusable traffic representations that can be adapted to new modulation categories with relatively few labeled examples.

This property helps explain the performance gap between BM-Net and the evaluated baselines. 
Most baseline models are optimized for passive website-fingerprinting settings and are trained directly on the available labeled target data. 
When the target dataset is small, these models have limited opportunity to learn stable representations of low-frequency bandwidth modulation, especially under Tor-induced jitter and path variability. 
In contrast, BM-Net benefits from pre-trained sequential representations before seeing the fine-grained labels, making it more data-efficient in the evaluated setting.

We emphasize that this result should be interpreted as evidence of data-efficient adaptation under our experimental conditions, not as a guarantee of identical performance across all Tor paths, modulation parameters, or defense configurations. 
Nevertheless, it suggests that separating self-supervised representation learning from supervised modulation classification is a useful design choice for traffic-analysis tasks where high-quality labeled traces are expensive to collect.

\subsection{Robustness and Limitations under Challenging Network Conditions}
\label{sec:sensitivity}

We further discuss robustness under two challenging conditions that are reflected in our dataset: cross-region Tor paths and client-side traffic-analysis defenses. 
The goal of this section is not to claim exhaustive robustness against all deployment scenarios, but to clarify that the reported results are obtained from traces that already include non-local network paths and defense-induced traffic variation.

\subsubsection{Cross-Region Network Variability}

A common concern for traffic analysis is whether performance measured under controlled or semi-controlled conditions transfers to long-haul Internet paths. 
Long-haul Tor paths may introduce latency variation, packet reordering, relay-load fluctuation, congestion, and circuit changes, all of which can distort the injected bandwidth pattern before it is observed at the exit side.

Our real-world dataset includes traces collected from geographically distributed Tor clients and exit relays. 
Therefore, the reported binary and fine-grained classification results are not obtained only from local-area or laboratory-only traffic. 
They include cross-region measurement noise caused by public Internet routing and Tor relay variability.

However, we do not claim that the same performance will hold for all possible long-haul paths or network states. 
More volatile routes, higher packet loss, stronger congestion, or frequent circuit replacement may reduce waveform separability and degrade fine-grained modulation classification. 
A more systematic evaluation across controlled RTT ranges, packet-loss levels, relay loads, and circuit lifetimes is left for future work.

\subsubsection{Defense-Aware Dataset Construction}

The dataset used in our evaluation includes traces collected under client-side traffic-analysis defenses, including WTF-PAD and Walkie-Talkie. 
These defenses were originally designed to reduce passive traffic-analysis leakage by injecting adaptive padding, reshaping bursts, or modifying timing patterns. 
Including such traces makes the evaluation more challenging than testing only on undefended Tor traffic, because the classifier must distinguish bandwidth-modulation patterns in the presence of defense-induced timing and burst-shape variation. This defense-aware construction makes the learning problem closer to a realistic deployment setting, where natural traffic variation and client-side obfuscation may coexist with active bandwidth perturbation.

The results in Section~\ref{sec:results} therefore reflect performance under a mixed setting that includes both natural Tor variability and defense-induced traffic transformations. 
This setting provides evidence that the evaluated active bandwidth perturbations can remain detectable even when passive-analysis defenses are present in the data distribution.

The reason is that WTF-PAD and Walkie-Talkie primarily operate on packet timing, padding, and burst-level structure, whereas NATA imposes a time-varying throughput constraint at the upstream shaper. 
Client-side padding and burst reshaping may alter the logical traffic pattern, but they do not directly remove the rate limit imposed by the upstream bottleneck. 
When the shaping rate decreases, excess traffic may be delayed, queued, or dropped depending on buffer state and transport-layer behavior. 
As a result, the exit-side throughput can still retain part of the imposed bandwidth-modulation pattern.

Nevertheless, this result should be interpreted carefully. 
Because defended traces are included in the training distribution, the experiment evaluates defense-aware detection rather than zero-shot generalization to unseen defenses. 
It does not rule out stronger or adaptive defenses specifically designed against active bandwidth perturbation, such as deliberate rate smoothing, randomized circuit rotation, pluggable transports, bridge usage, or client-side detection of rhythmic low-frequency throughput variation.

In summary, the evaluation instantiates the theoretical model from Section~\ref{sec:theoretical_modeling} with simulated exit-observation probabilities and real-world classification measurements collected under mixed network and defense conditions. 
Under the evaluated settings, BM-Net achieves strong binary detection and fine-grained modulation classification performance on the collected traces. 
Taken together, these results suggest that active bandwidth perturbation can increase traffic-correlation risk under the considered infrastructure-level adversary model, while leaving open the need for broader evaluation against adaptive defenses and more diverse network conditions.

\section{Discussion}
\label{sec:discussion}

This section discusses limitations, operational considerations, defense implications, and ethical safeguards. 
Our results should be interpreted under the threat model and experimental settings defined in this paper, rather than as evidence that all Tor traffic is equally vulnerable to active bandwidth perturbation.

\subsection{Limitations and Circuit Dynamics}

\paragraph{Circuit stability and rate selection}
Active bandwidth perturbation interacts with Tor circuit dynamics and transport-layer behavior. 
Overly aggressive throttling may cause stalls, retransmissions, timeouts, or circuit replacement, reducing usable evidence for correlation. 
Thus, the shaping rate must balance exit-side detectability against connection stability. 
In our experiments, the low-rate phase is selected empirically rather than treated as a universal Tor keep-alive threshold. 
The usable range may vary with relay load, path length, TCP congestion control, Tor multiplexing, website behavior, and background network conditions.

\paragraph{User-perceived degradation}
The low-rate value is applied as the transient trough of a time-varying waveform rather than as a continuous flat throttle. 
Thus, it may appear as short-lived throughput fluctuation, temporary jitter, or a brief page-load stall rather than persistent degradation. 
However, if the low-rate phase is too long, frequent, or aggressive, it may noticeably degrade browsing performance and trigger retries, stalls, or circuit replacement. 
Practical deployment is therefore constrained by a detectability--usability trade-off, and systematic user-experience measurements remain future work.

\paragraph{Circuit reselection}
Tor's circuit management can implicitly mitigate aggressive perturbation: repeated stalls or poor throughput may cause a connection to fail or be replaced before sufficient evidence is collected. 
This creates a detectability--stability trade-off, where stronger perturbations are easier to detect but more disruptive.

\subsection{Operational Considerations}

Real-world congestion, relay load, cross-traffic, routing changes, and diurnal variation can distort or obscure the injected pattern. 
Our cross-region measurements suggest that the evaluated modulation patterns can remain detectable in the collected non-local traces, but performance may degrade under more volatile paths or transport conditions. 
Repeated observations may increase correlation probability under the simplifying independence assumptions in Section~\ref{sec:theoretical_modeling}, although correlations across windows may reduce independent evidence.

At the entry side, bandwidth perturbation can be implemented using token-bucket queueing in Linux \texttt{tc}, avoiding application-layer parsing and payload decryption. 
However, per-flow shaping still requires Tor-connection identification, queue management, rate-state maintenance, and buffering. 
The overhead is likely manageable for targeted flows, but should not be assumed to scale to arbitrary ISP-wide deployment without further measurement. 
Bridges, pluggable transports, VPNs, or other tunneling settings may also complicate Tor-connection identification.

\subsection{Defense Implications}
\label{subsec:defense_dilemma}

Our evaluation suggests that defenses designed primarily for passive traffic analysis may not fully address active bandwidth perturbation. 
Padding and burst-shaping can modify packet timing and traffic-volume patterns, but they do not directly remove an upstream rate limit imposed by an adversary-controlled shaper. 
Potential defenses include detecting rhythmic low-frequency throughput variation, rotating circuits under suspicious rate patterns, using bridges or pluggable transports, applying adaptive rate smoothing, or designing network-layer mechanisms that reduce the visibility of imposed bandwidth constraints. 
Each defense introduces trade-offs among bandwidth overhead, latency, deployability, and robustness.

\subsection{Ethical Considerations}

Because this work involves active bandwidth manipulation, we followed safeguards to limit collateral impact. 
All client-side traffic was generated by the authors, and shaping was applied only to our experimental traffic. 
We did not intentionally perturb unrelated Tor users, inspect application-layer content, or decrypt payloads. 
Exit-side measurements were filtered to experimental traces, and perturbations were temporary and limited to the controlled measurement campaign.

\section{Conclusion}
\label{sec:conclusion}

In this paper, we studied Non-invasive Active Traffic-correlation Analysis (NATA), an infrastructure-level traffic-correlation setting in which an adversary introduces controlled bandwidth perturbations into selected Tor-related traffic. 
Unlike passive traffic analysis, which relies only on naturally occurring timing and volume patterns, NATA uses time-varying throughput constraints to create measurable bandwidth signatures. 
The attack does not require endpoint compromise, Tor-browser modification, packet-payload decryption, or application-layer content inspection.

We developed a probabilistic model that separates traffic-correlation risk into three components: exit-relay observation probability, perturbation detection probability, and modulation classification probability. 
The model captures how monitoring multiple flows or repeated observation windows can increase the overall correlation probability, while also making clear that such aggregation relies on simplifying independence assumptions and exhibits diminishing marginal gains. 
Using \textit{tornettools}-based scaled simulations, we estimated the bandwidth-weighted probability that watermarked traffic is observed by adversary-controlled exit relays.

We also designed BM-Net, a selective state-space neural framework for detecting and classifying bandwidth-modulation patterns from noisy Tor traffic. 
BM-Net uses self-supervised masked pre-training followed by supervised fine-tuning, reducing the amount of fine-grained labeled data required for modulation classification. 
In our real-world Tor measurements, BM-Net achieves a 99.65\% binary detection F1 score and a 97.5\% macro-F1 score for fine-grained modulation classification under the evaluated settings.

Overall, our results suggest that active bandwidth perturbation can increase traffic-correlation risk under a clearly defined infrastructure-level adversary model. 
These findings highlight the need to consider bandwidth-control side channels when designing and evaluating defenses for low-latency anonymity networks. 
At the same time, the attack is constrained by circuit stability, user-visible performance degradation, traffic identification assumptions, and deployment overhead. 
Future work should further evaluate user-experience impact, circuit-reselection behavior, adaptive defenses, and larger-scale network conditions.

\end{document}